\documentclass{llncs}

\usepackage[colorinlistoftodos]{todonotes} % TODO notes

\usepackage[caption=false]{subfig} % Subfigures

\usepackage{enumitem}
\newlist{inparaenum}{enumerate}{2}% allow two levels of nesting in an enumerate-like environment
\setlist[inparaenum]{nosep}% compact spacing for all nesting levels
\setlist[inparaenum,1]{label=\bfseries\arabic*.}% labels for top level
\setlist[inparaenum,2]{label=\arabic{inparaenumi}\emph{\alph*})}% labels for second level

\begin{document}

\title{Community Targeted Phishing}

\subtitle{A Middle Ground Between Massive and Spear\\ Phishing through Natural Language Generation}

\titlerunning{Community Targeted Phishing through NLG}  % abbreviated title (for running head)
%                                     also used for the TOC unless
%                                     \toctitle is used
%
\author{Alberto Giaretta\inst{1} \and Nicola Dragoni\inst{1}\inst{2}}
\authorrunning{Alberto Giaretta et al.} % abbreviated author list (for running head)
%
%%%% list of authors for the TOC (use if author list has to be modified)
\tocauthor{Alberto Giaretta, Nicola Dragoni}
\institute{Centre for Applied Autonomous Sensor Systems (AASS),\\ \"Orebro University, Sweden,
\email{alberto.giaretta@oru.se},%\\ WWW home page:
%\texttt{http://users/\homedir iekeland/web/welcome.html}
\and
DTU Compute, Technical University of Denmark,\\ Denmark
\email{ndra@dtu.dk}
}

\maketitle              % typeset the title of the contribution

\begin{abstract}
Looking at today phishing panorama, we are able to identify two diametrically opposed approaches. On the one hand, massive phishing targets as many people as possible with generic and preformed texts. On the other hand, spear phishing targets high-value victims with hand-crafted emails. While nowadays these two worlds partially intersect, we envision a future where Natural Language Generation (NLG) techniques will enable attackers to target populous communities with machine-tailored emails. In this paper, we introduce what we call Community Targeted Phishing (CTP), alongside with some workflows that exhibit how NLG techniques can craft such emails. Furthermore, we show how Advanced NLG techniques could provide phishers new powerful tools to bring up to the surface new information from complex data-sets, and use such information to threaten victims' private data.
\end{abstract}

\section{Introduction}\label{sec:intro}
Even though many years have passed since the Internet has come into our lives, some of its atavistic problems still have to be addressed. Phishing is one of them, and we foresee that  its severity could even worsen in the coming years. 

On the one hand, we have a massive phishing approach where attackers fill simple templates with some basic information about the target, aiming to create simple and sound emails with the minimum effort possible.

On the other hand, we have \textit{spear phishing} emails where attackers experienced in social-engineering techniques write articulate and specific emails. Even though these emails are carefully forged and hard to detect, the creation process requires such an effort that they are reserved only for high-level targets. Spear phishing techniques are more and more popular into hackers' toolboxes, as an example they are essential to strike \textit{Advanced Persistence Threats} (\textit{APTs})~\cite{6231617}.

\begin{figure}[th]
\center
\includegraphics[trim=0mm 0mm 0mm 0mm,clip,width=0.80\textwidth]{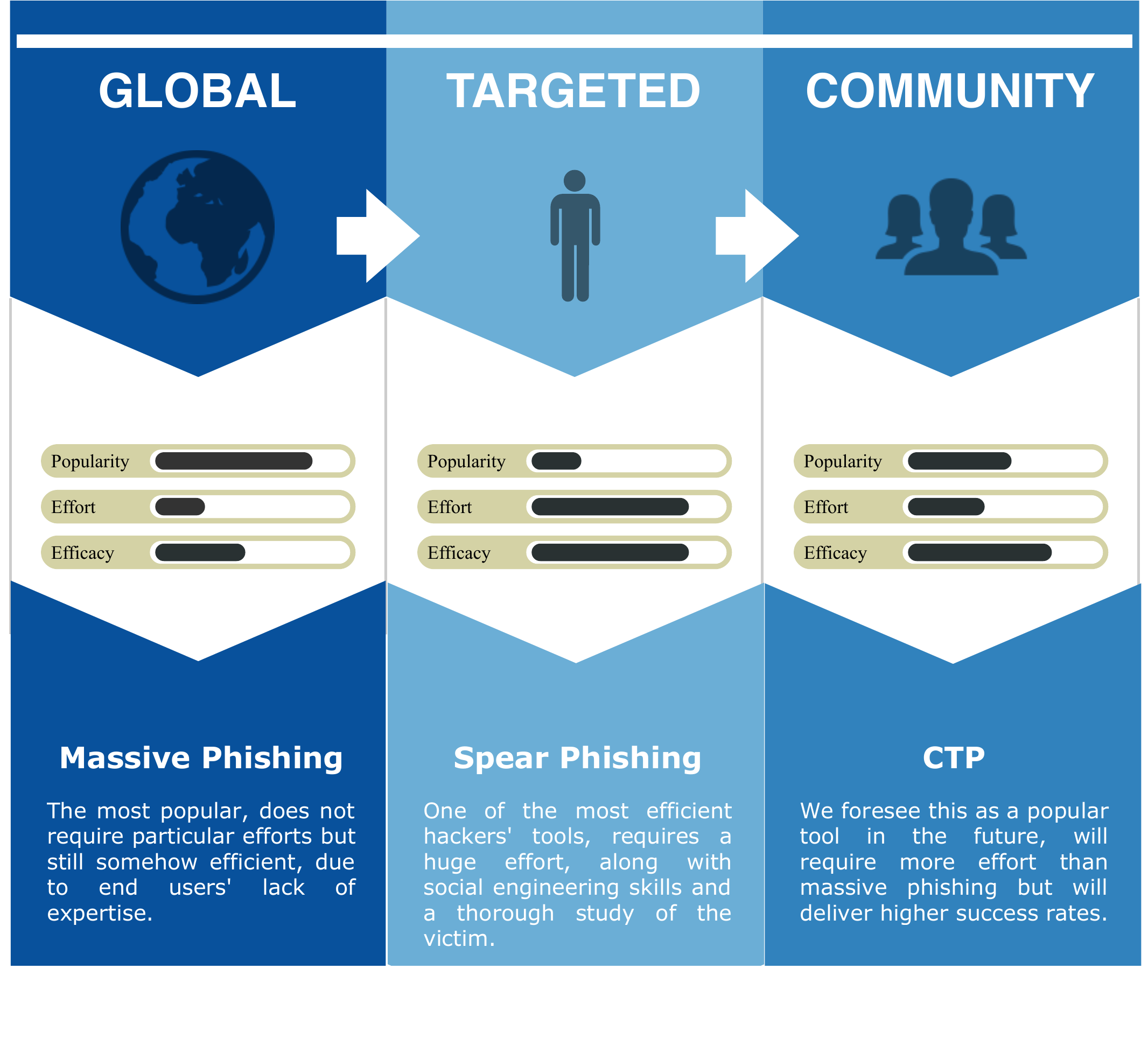}
\caption{Phishing naturally evolved from a massive approach to a manually hand-crafted tool, used by hackers to target high value victims. We foresee a midway approach in the future, which combines effectiveness and cheapness. Value bars present an estimate of each characteristic, not actual statistic data.}
\label{fig:spam_evolution}
\end{figure}

Although they have the same goal, until today these two worlds have been somehow distinct from each other: massive production on one side, craftsmanship on the other one. We envision a future where a middle ground will be prominent, thanks to sophisticated production processes that go under the name of \textit{Natural Language Generation} techniques, which will enable to target specific categories with machine-tailored messages. We call this new approach to phishing \textit{Community Targeted Phishing} (\textit{CTP}), and we show in Figure~\ref{fig:spam_evolution} an example of this evolution, as we foresee it.

To the best of our knowledge, at the time of writing only one recent work about NLG applied to emails generation has been published~\cite{Baki:2017:SEE:3052973.3053037}. In that work, authors show that NLG-based emails can be extremely effective to deceive people. Apart from such work, no other studies dig into how NLG fits into the big picture, nor how smart partitioning the targets could dangerously combine with NLG techniques. Strictly related, Natural Language Processing (NLP) goes in the opposite direction of NLG by extracting information from written texts, and it has been proposed as a solution to develop spam filters~\cite{Verma2012}.

This paper is organized as follows: Section~\ref{sec:nlg} gives a basic overview about NLG and its core concepts, useful to read through this paper. Section~\ref{sec:ctp} digs into our predictions introducing our CTP concept with related workflows, in order to clarify how Template-Filling NLG could enable attackers to target the victims in the future. Then, Section~\ref{sec:adv_nlg} goes even further, making some hypothesis about how Advanced NLG techniques could enable attackers to gain victims' trust and increase the success rate of their phishing attempts. Finally, Section~\ref{sec:conc} wraps up our conclusive thoughts and lists some directions that we would like to take in the future.

\section{Natural Language Generation (NLG)}\label{sec:nlg}
First of all, what do we talk about when we talk about Natural Language Generation? A simple definition is: ``Natural language generation (NLG) systems generate texts in English and other human languages''~\cite{Reiter2010}, yet this task is easier said than done. As a matter of fact, NLG systems have to make quite a number of different choices, in order to achieve the most suitable text for the specific purpose. As an example, psycho-linguistic models of language comprehension show that reflective pronouns usually improve a text readability and thus should be normally used, but such pronouns should be usually avoided in safety-critical texts, like operation manuals for nuclear power plants~\cite{Reiter2010}.

From a high-level perspective, NLG is traditionally divided in three macro-categories, namely \textit{Canned Text}, \textit{Template-Driven (or Template Filling)}, and \textit{Advanced (or Proper) NLG}. Canned Text is the most basic approach, where a text is pre-generated (usually by a human being) and is used when the right moment comes. Even though different texts can be merged with some \textit{glue texts}, this technique is hardly flexible and unable to adapt even to little domain variations. As previously pointed out~\cite{VanDeemter:2005:RVT:1122624.1122626}, ``Canned text is the borderline case of a template without gaps''. 

Talking about Template-Driven, this approach consists into defining some templates that exhibit \textit{gaps}, to be filled with the correct information gathered from an external source, such as a database. The third approach, the Advanced NLG, is sometimes called proper NLG to highlight the fact that a NLG system implies a large variety of choices, and the two aforementioned approaches oversimplify the matter by using preformed texts. In particular, some authors argue that Template-Driven approaches are not proper NLG and tend to dismiss them, while others affirm that Template-Driven systems have been so much refined through the years that the distinction between them and pure NLG is too blurred to hold, nowadays~\cite{VanDeemter:2005:RVT:1122624.1122626}. 

No matter what kind of approach is chosen, even though NLG systems are far from being widespread, they have been used in many applications. One of the classic examples are weather forecasts, which are automatically generated from meaningful database entries; other fields implemented NLG in their routines, such as soccer, financial, technical, and scientific automatic reports.

\section{Aiming at Groups: Community Targeted Phishing (CTP)}\label{sec:ctp}
What massive phishing lacks is a deeper comprehension about recipients' relationships, characteristic which is totally neglected nowadays. Yet, dividing users in meaningful subsets could help phishers to impersonate some group members, and attack the others, in a much more effective way. This is not done yet, mostly because traditional phishing is still effective, but spam filters are proving themselves more and more reliable~\cite{6489877}~\cite{Blanzieri2008}. One popular approach to develop such filters, is to apply the Bayesian filtering. In order to discern a legitimate email from a malicious one, this process checks every single word against a table, which contains the most commonly used words in spam emails. Even though advanced adversarial machine learning techniques can be used to poison the whole process~\cite{Nelson:2008:EML:1387709.1387716}, this approach is effective.

Such filters require training and leverage the strong points of phishing, which are massiveness and repetitiveness, to get statistically meaningful data. If phishers could smartly divide this huge user blob in smart subsets, they could forge more complex emails and deprive Bayesian spam filters of the aforementioned vital statistical data. We call this approach Community Targeted Phishing (CTP), and in the following subsection we provide some examples on how it could be applied.

\subsection{Case Study: Scientific Community}
One community which we are familiar with is the scientific one. As a matter of fact, researchers working in the same research field mostly share many interests and know quite a number of colleagues. Furthermore, the email exchange for collaborations, sharing of ideas and call for papers are the routine. On the average, if computer science researchers are enough knowledgeable about security and spam, members of other scientific fields usually exhibit very basic computer skills, which makes them easy targets for malicious attackers.

Scientists already receive a lot of scam emails from shady conferences and journals, which name resembles a legitimate venue and invite the recipients to submit their work. Sometimes people willingly submit to pay-for-publish venues, some other times they are fooled into believing that they are actually submitting to an honest venue. But these phishing emails are really simple in their form, as attackers just fill the first line with something along the line of ``Dear Prof. Smith'' without any kind of variation in the preformed text and without trying to impersonate an actual researcher. Furthermore, in many cases the attackers even fail to parse correctly the title, as it often happens for PhD Students to be addressed as Dr., Prof., or similarly wrong titles. Succeeding into impersonate a real person through an automated process, on the other hand, is a far more serious (and hard to achieve) threat and this is what we address in our paper.

As a case study, we propose a mechanism that fills two proposed templates by utilizing Template-Driven NLG techniques, and the related workflows that show how the whole email forging process could take place. As a matter of fact, we believe that even such basic templates could successfully fool a good number of people. The main problem that we see with this approach is that it still requires some human effort, to understand how a community usually interacts and what subsets should be excluded or included into the attack. As an example, in Subsection~\ref{subsec:prop_txt_2} we will see that targeting very close colleagues with that particular template should be avoided.

\subsubsection{Template 1}
\begin{figure}[th]
\center
\includegraphics[trim=0mm 0mm 0mm 0mm,clip,width=0.60\textwidth]{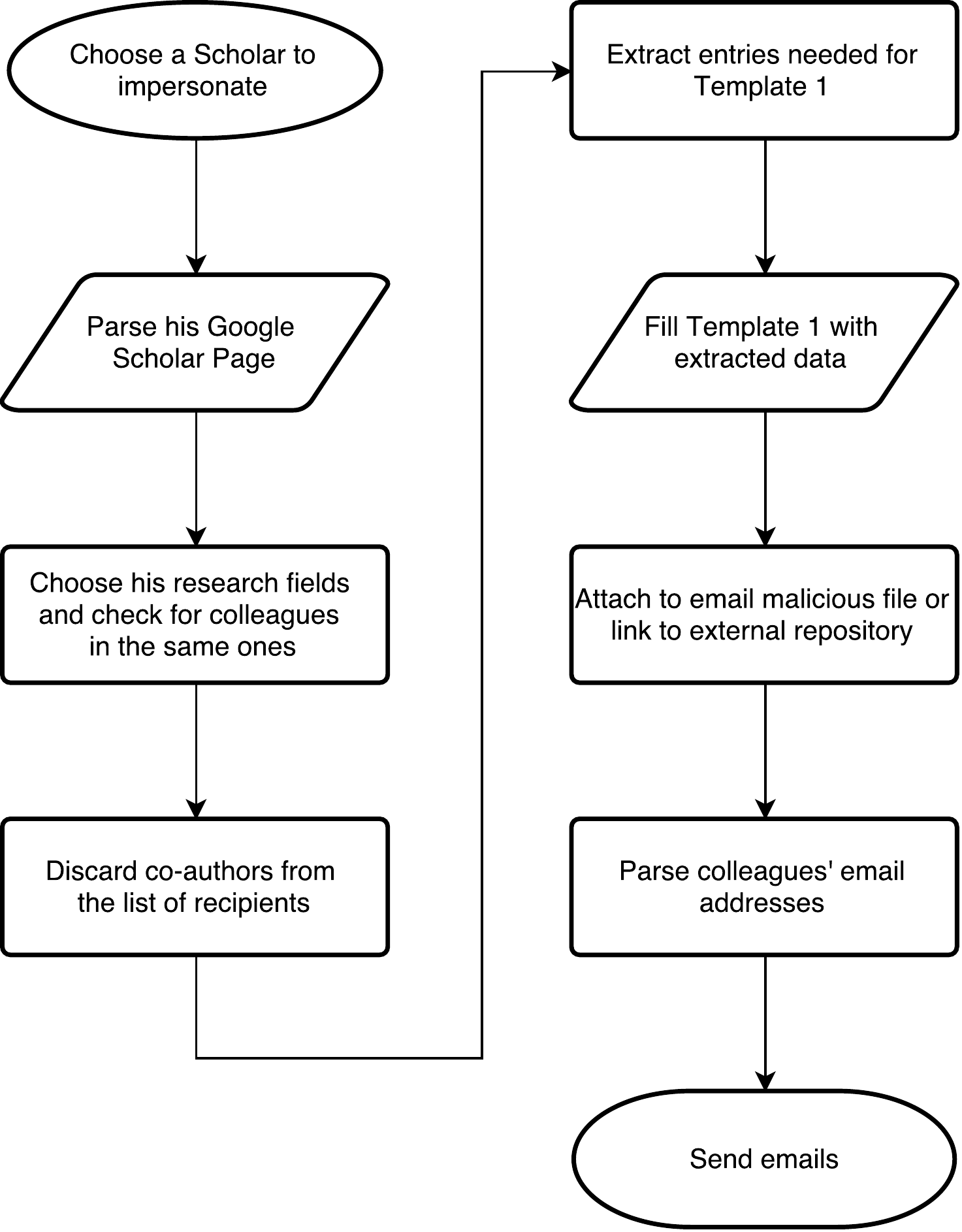}
\caption{Workflow of Template-driven NLG applied to Template 1.}
\label{fig:workflow_1}
\end{figure}

The first proposal takes advantage of the fact that researchers know many names in their field, but the relationships tend to be very weak. Yet, networking is an important part in scientists' lives, therefore a proposal from a respected name in the field could be an easy lure for a young researcher.
The template is: ``Dear [Colleague Name], I've read your recent work entitled [Manuscript Title] and I found it quite interesting. I've come up with some ideas about that same topic and I would be enthusiastic to work with you on this. At [Fake Google Drive Link] you will find a brief recap about my insights. Hope to hear from you soon, best regards [Scholar Name]".  As before, the link will redirect the user to a fake login page which resembles Google login page. Figure~\ref{fig:workflow_1} shows the workflow to produce the proposed text.

\subsubsection{Template 2}\label{subsec:prop_txt_2}
\begin{figure}[th]
\center
\includegraphics[trim=0mm 0mm 0mm 0mm,clip,width=0.60\textwidth]{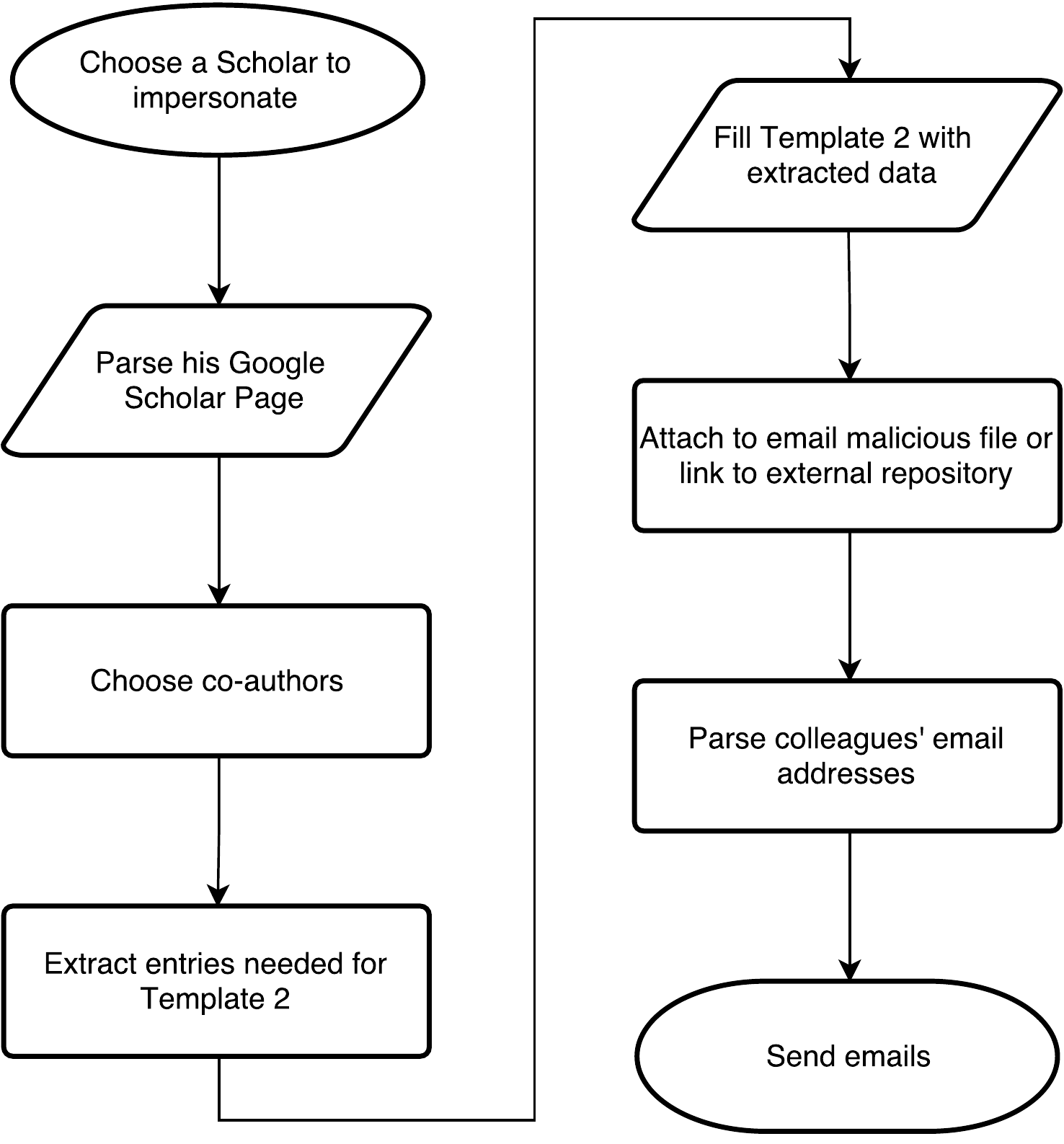}
\caption{Workflow of Template-driven NLG applied to Template 2.}
\label{fig:workflow_2}
\end{figure}

The second approach is riskier because the attacker has to impersonate someone close to the victim. A forged email could sound way too formal (e.g., nicknames are normally used between acquaintances), but it could pay since colleagues often read privately each others' works, as a form of preliminary peer-review.
The template is: ``Hi [Colleague Name], as you know I'm working on [Topic Name] and I've written the attached paper. I would like to have some suggestions from you about it, since I'm a little uncertain about the solidity of the whole work. At [Fake Google Drive Link] you will find my draft. Thanks so much in advance, see you soon [Scholar Name]''. As before, the link will redirect the user to a fake login page which resembles Google login page. Figure~\ref{fig:workflow_2} shows the workflow to produce the proposed text.

%\begin{figure}[ht]
%\centering
%\subfloat[Template 1]{\includegraphics[width=0.45\textwidth]{images/NLG_Template-Driven_-_Workflow_1.pdf}
%\label{fig:workflow_1}}
%\hspace{1em}
%\subfloat[Template 2]{\includegraphics[width=0.45\textwidth]{images/NLG_Template-Driven_-_Workflow_2.pdf}
%\label{fig:workflow_2}}
%\caption{Workflow that shows how Template-driven NLG could be used with the proposed templates} %$\frac{1}{1+(\frac{x}{30})^4}$
%\end{figure}

\section{Advanced NLG}\label{sec:adv_nlg}
In this section, we take one step more in our predictions and depict a scenario where Advanced NLG is used to create more convincing phishing emails.

As an example, an attacker could parse Google Scholar and look for researchers' h-index trend during the past 5 years and their performance with respect to their colleagues in the same field. With such data, the attacker could use NLG techniques to summarize the trends in a natural language and incorporate this new information in a phishing email. The result would be something along the line of ``Dear [Target Name], I am Prof. [Reputable Name] from [Reputable University], we are currently expanding our lab and we are evaluating some possible candidates. I have personally checked your Google Scholar account and noticed that you h-index is high with respect of the average in your field, and considerably grew over the last 5-years span. Therefore, we would like to propose you a position. You can enrol at the following link: [Phishing Link]''. If the victim is successfully tricked, they might fill the recruitment form and give up valuable information to the attacker.

Besides the research community, similar techniques could be applied to social network users. For example, an attacker could use NLG techniques to identify and summarize a specific interest of a social network account and forward a private message along the line of: ``Hello [Victim], I've seen that you're really into [Music Genre]. I'm an event organizer and I would like to inform you that the next weekend, at the [Big Square] in [Victim Hometown], [Good Music Player] will give a short exhibition. All the details in the following link, feel free to attend! [Phishing Link]''. The phishing link would be a fake login page which mimics the social network one, created by the attacker. If the victim believes that a disconnection due to technical issues happened, they might try to login again, which would result in stolen credentials.

Despite the very simplistic nature of the aforementioned examples, the capability of translating heterogeneous information into natural language text can play a huge part in the next future of automatic phishing generation. Indeed, through Advanced NLG techniques attackers do much more than creating automatic text: they pinpoint submerged information which would take a huge effort to manually discover.

\section{Conclusion}\label{sec:conc}
In this paper, firstly we have given a brief overview about the concept of Natural Language Generation (NLG) techniques and how these could be used to forge phishing emails. Moreover, we have expressed our concerns about how NLG could help hackers to easily target a community, such as the scientific one, proposing the concept of Community Targeted Phishing (CTP). We have provided two workflows that show how a Template-driven approach could be used to create phishing messages, and then we analysed how Advanced NLG techniques could enable phishers to achieve complex (thus, harder to spot) emails. 

Future works will investigate different NLG techniques, in order to understand which ones provide the best results in terms of credibility and effectiveness. Moreover, we would like to investigate the variety that Advanced NLG techniques can ensure in text generation, as suggested in Section~\ref{sec:adv_nlg}, and evaluate how traditional Bayesian spam filters perform against such phishing texts.

This work is just a first step, useful to introduce the very concept of CTP and raise awareness about a problem that might soon show up. We believe that using a NLG approach to target people with similar interests could be worthwhile, since it would allow to create emails more effective than the general ones, and cheaper than the spear phishing ones.

% TODO list
%\listoftodos

%
% ---- Bibliography ----
%

\bibliographystyle{splncs03}
\bibliography{bibliography}

\end{document}